\documentclass[twocolumn]{aastex631}
\usepackage{soul}

\usepackage{enumitem}
\begin{document}

\title{An upgraded GMRT and MeerKAT study of radio relics in the low mass merging cluster \\PSZ2 G200.95-28.16}

\author[0009-0007-8409-4233]{Arpan Pal}
\affiliation{National Centre for Radio Astrophysics, Tata Institute of Fundamental Research, S. P. Pune University Campus, Ganeshkhind, Pune, 411007}

\author[0000-0003-1449-3718]{Ruta Kale}
\affiliation{National Centre for Radio Astrophysics, Tata Institute of Fundamental Research, S. P. Pune University Campus, Ganeshkhind, Pune, 411007}

\author[0000-0001-8468-9164]{Qian H. S. Wang}
\affiliation{Department of Physics \& Astronomy, The University of Utah, 115 South 1400 East, Salt Lake City, UT 84112, USA}

\author[0000-0001-9110-2245]{Daniel R. Wik}
\affiliation{Department of Physics \& Astronomy, The University of Utah, 115 South 1400 East, Salt Lake City, UT 84112, USA}

\begin{abstract}
Diffuse radio sources known as radio relics are direct tracers of shocks in the outskirts of merging galaxy clusters. PSZ2 G200.95-28.16, a low-mass merging cluster($\textrm{M}_{500} = (2.7 \pm 0.2) \times 10^{14}~\mathrm{M}_{\odot}$) features a prominent radio relic, first identified by \citealp{2017MNRAS.472..940K}. We name this relic as the Seahorse. The MeerKAT Galaxy Cluster Legacy Survey has confirmed two additional radio relics, R2 and R3 in this cluster. We present new observations of this cluster with the Upgraded GMRT at 400 and 650 MHz paired with the Chandra X-ray data.  
The largest linear sizes for the three relics are~1.53 Mpc, 1.12~kpc, and 340~kpc. All three radio relics are polarized at 1283~MHz.  Assuming the diffusive shock acceleration model, the spectral indices of the relics imply shock Mach Numbers of $3.1 \pm 0.8$ and $2.8 \pm 0.9$ for the Seahorse and R2, respectively. 
The Chandra X-ray surface brightness map shows two prominent subclusters, but the relics are not perpendicular to the likely merger axis as typically observed; no shocks are detected at the locations of the relics.
We discuss the possible merger scenarios in light of the low mass of the cluster and the radio and X-ray properties of the relics. The relic R2 follows the correlation known in the radio relic power and cluster mass plane, but the Seahorse and R3 relics are outliers. 
We have also discovered a radio ring in our 650~MHz uGMRT image that could be an Odd radio circle candidate. 
\end{abstract}
\keywords{Galaxy clusters, Radio relics, Low-mass cluster merger, Non-thermal emission, Individual galaxy cluster: PSZ2 G200.95-28.16}
\section{Introduction} \label{sec:intro} 
Diffuse radio emission associated with the intra-cluster medium (ICM) of merging galaxy clusters can be broadly classified into two main categories: radio halos and radio relics (For review; \citealp{2019SSRv..215...16V}). Radio halos are diffuse radio sources found at the cluster centres with surface brightness typically following the thermal ICM distribution \citep{https://www.aanda.org/articles/aa/pdf/2001/14/aah2391.pdf, https://doi.org/10.1051/0004-6361:20010581, https://www.aanda.org/articles/aa/pdf/2005/36/aa3016-05.pdf}. On the other hand, radio relics are arc-like, diffuse polarized sources (\citealp{https://arxiv.org/abs/astro-ph/9712293, 2009A&A...494..429B, https://arxiv.org/abs/1206.6102}), typically found in the cluster outskirts (\citealp{1991A&A...252..528G}; \citealp{https://arxiv.org/abs/1010.4306}). When two galaxy clusters merge, the merger-induced shocks and injected turbulence amplify the ICM magnetic fields and (re)accelerate the cosmic ray electrons via mechanisms like diffusive shock acceleration (DSA) (\citealp{1977DoSSR.234.1306K};\citealp{1977ICRC...11..132A, 1978MNRAS.182..147B, 1987PhR...154....1B}). 
The (re)accelerated electrons in the presence of cluster magnetic fields will emit synchrotron radiation in the form of radio relics. 
\begin{figure*}[ht!]
\epsscale{1.2}
\plotone{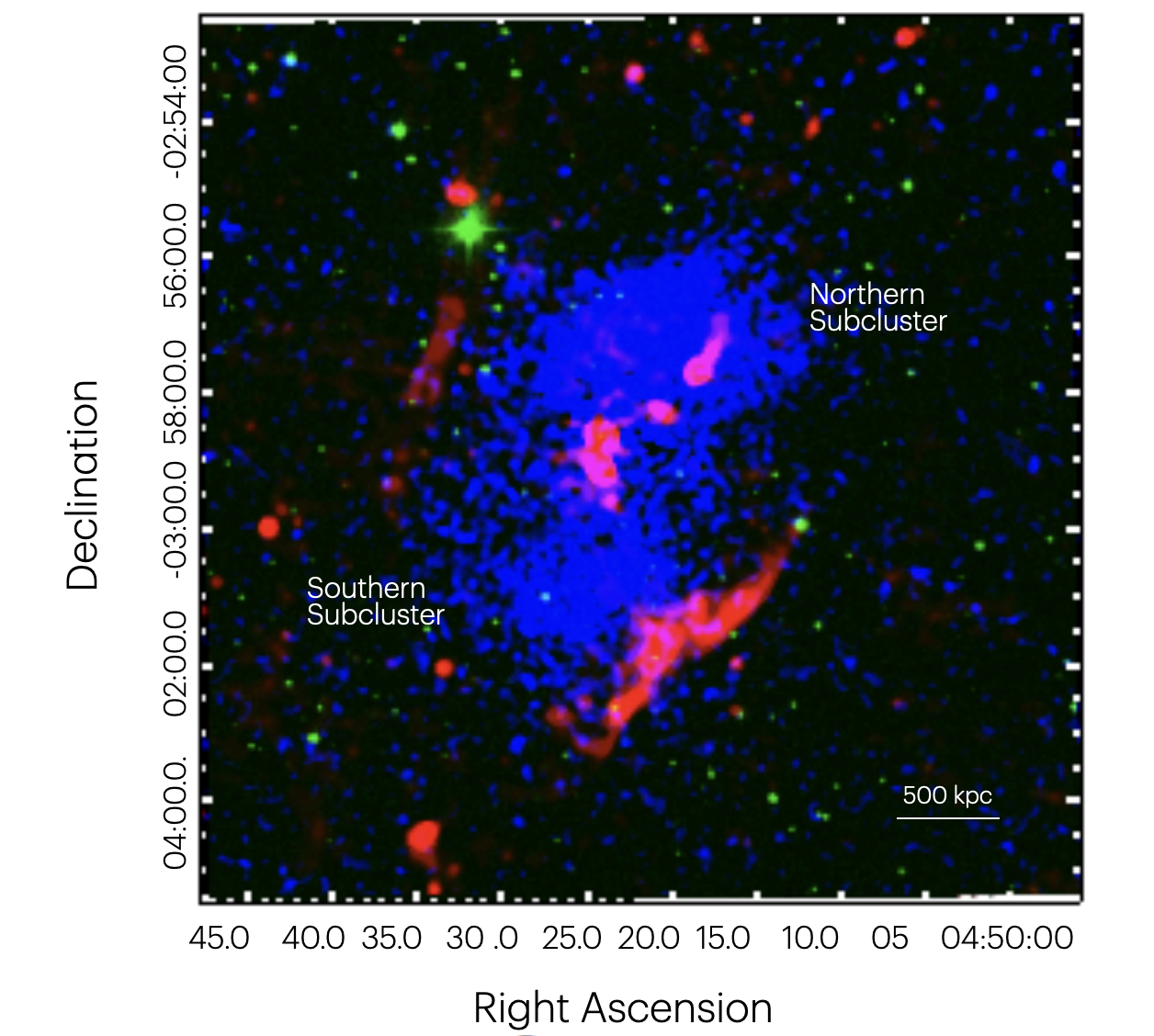}
\label{rgb}
\caption{X-ray, optical and radio composite image of the PSZ2 G200.95-28.16 field with uGMRT 650 MHz smoothed $10\arcsec \times 10\arcsec$ image in red, DSS-2 R-band image in green, and Chandra $0.8-4$ keV X-ray image in blue (smoothed with a $10\arcsec$ Gaussian kernel).}
\end{figure*}

The radio relics show a gradual spectral steepening from outer edge to inner edge \citep[e.g.,][]{https://doi.org/10.1093/mnras/stu1658, 2017MNRAS.472..940K, 2020A&A...636A..30R}. This is the result of the ageing of relativistic electrons via inverse Compton and synchrotron losses in the wake of the shocks. The spectral behaviour of integrated flux densities is usually described by a power-law \citep{2020A&A...642L..13R,galaxies9040111}. In polarization studies, radio relics typically exhibit an average polarization fraction of 15-20\%  (e.g. \citealp{2010Sci...330..347V, 2012MNRAS.426.1204K,2022Galax..10...10H}) at GHz frequencies, with the electric field position angles (EVPA) typically aligned with the shock normal(e.g. \citealp{2010Sci...330..347V, 2021ApJ...911....3D}). This spatio-spectral trend, the observed power-law spectral indices, and the existence of polarization serve as evidence in favour of DSA to be the acceleration mechanism. However, the acceleration efficiency of DSA is found not to be adequate to produce the observed radio power for some radio relics, assuming the electrons are directly accelerated from the thermal ICM pool \citep[]{10.1088/0004-637X/734/1/18,2013A&A...555A.110S,2020A&A...634A..64B}. A population of pre-existing supra-thermal or mildly relativistic seed of electrons from radio galaxies and active galactic nuclei (AGN) activity could be a possible solution \citep{1998astro.ph..5367E,2005ApJ...627..733M,2011ApJ...734...18K}. Current frameworks are still to resolve the issues with acceleration inefficiency and the origin of cosmic ray electrons and ICM magnetic fields.

Among the clusters hosting radio relics, a special class is of those hosting double radio relics. In these cases, the merger axis is nearly in the plane of the sky. This favourable geometry can be used to study both the shocks and can constrain the cluster merger parameters robustly due to reduced uncertainties of projection effects. Clusters with two radio relics are relatively rare as specific merging configurations are needed to produce double radio relic systems. Clusters hosting more than two radio relics are even rarer compared to double radio relic systems. To date, only 4 systems (Abell 2744 \citep{2017ApJ...845...81P}; Sausage \citep{2017MNRAS.471.1107H}; 1RXS J0603.3+421 \citep{2012MNRAS.425L..76B} and Abell 746 \citep{2023arXiv230901716R}) are discovered to host multiple radio relics.
For example, the triple radio relic system in 1RXS J0603.3+421 has been interpreted as the result of a triple cluster merger \citep{2012MNRAS.425L..76B}. On the other hand, Abell 746, which hosts 4 radio relics, is thought to originate in multiple mergers, evident by the detection of X-ray shocks towards different axes \citep{2023arXiv230901716R}. The rarity of systems hosting multiple radio relics suggests the complexity of cluster mergers, likely involving interactions between more than two subclusters.

PSZ2 G200.95-28.16 is a low-mass merging cluster that hosts three radio relics. We have investigated this highly disturbed merging system at radio wavelengths using the upgraded Giant Metrewave Radio Telescope (uGMRT) at 400 and 650 MHz, as well as MeerKAT at 1283 MHz, and at X-ray wavelengths with the Chandra X-ray Observatory. The paper is organised as follows: In Sec.~\ref{sec:G200} we introduce the cluster and in Sec.~\ref{sec:obs}, radio observations and data analysis is summarised. X-ray data analysis is presented in Sec.~\ref{sec:X-ray}. The radio images, spectral and polarization results are presented in Sec.~\ref{sec:res}. We discuss the radio power and host cluster mass scaling relation, the shocks underlying the relics and the implied merging scenarios in Sec.~\ref{sec:discussion}. The conclusions are provided in Sec.~\ref{sec:con}. 
\begin{figure*}[ht!]
\epsscale{1.2}
\plotone{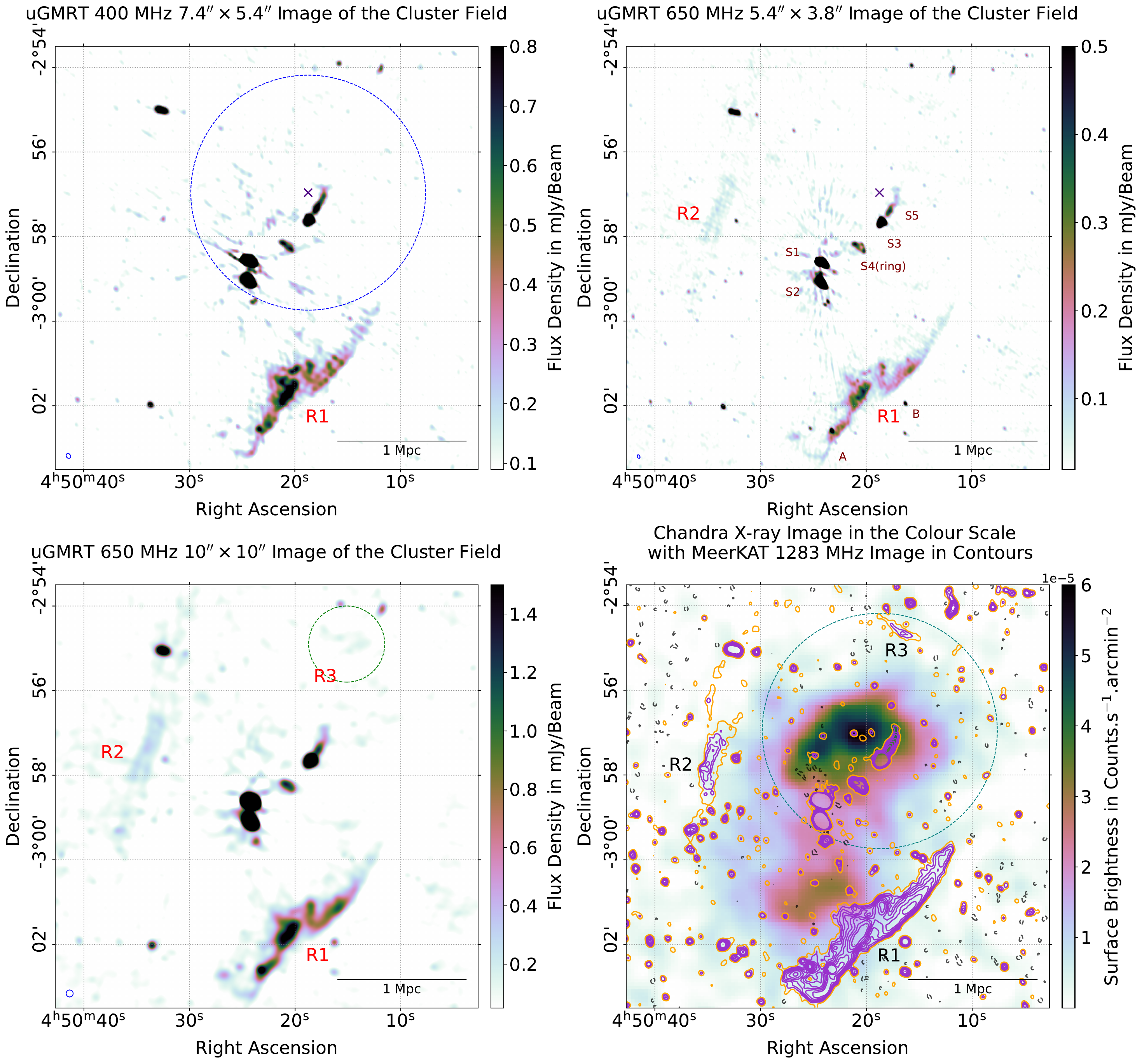}
\caption{
All the radio relics where they are detected are labelled with R1 (Seahorse), R2, and R3. Top left: uGMRT 400 MHz $7.4 \arcsec \times 5.7 \arcsec$ image of the PSZ2 G200.95-28.16. The blue `x' marks the location of the X-ray peak. The blue dotted circle signifies the region under $\mathrm{R}_{500}$ centred at the X-ray peak. Top right: uGMRT 650 MHz $5.4 \arcsec \times 3.8 \arcsec$ image of the cluster PSZ2 G200.95-28.16. All the relevant point sources are marked with S1-S5, A, and B. Bottom left: Smoothed uGMRT 650 MHz $10 \arcsec \times 10 \arcsec$ image of the field. To highlight the faint detection of the radio relic R3, we have marked a circle around it. Bottom right: $7.9 \arcsec \times 7.3 \arcsec$ MeerKAT 1283 MHz contours starting at $3\sigma$ with a multiplicative increment of $\sqrt{2}$ in magenta overlayed on Chandra 0.8-4 keV surface brightness image in colour scale. The orange contours represent 2$\sigma$ levels. The circle from the top left panel is shown in green.}
\label{field}
\end{figure*}

\begin{table*}
\centering
\caption{Summary of observations with uGMRT for PSZ2 G200.95-28.16.}
\begin{tabular}{ccccccccc}
\hline
\hline
Proposal& Frequency & Observing & On Soure & Bandwidth & Channels & Synthesized & Position &Sensitivity  \\
Code & & Date & Time &  &  & Beam & Angle & \\
 \hline
$38\_030$ & 400 MHz & 2020 July 12 & 5.0 Hours & 200 MHz & 2048 & $7.4 \arcsec \times 5.7 \arcsec$ & $33^{\circ}$ &50 $\mu$Jy/beam \\
ddtc286& 400 MHz & 2023 July 23 & 6.5 Hours & 200 MHz & 2048 & $7.4 \arcsec \times 5.7 \arcsec$ & $38^{\circ}$&40 $\mu$Jy/beam \\
$38\_030$& 650 MHz & 2020 July 13 & 5.0 Hours & 200 MHz & 2048 & $5.4 \arcsec \times 3.8 \arcsec$ &  $27^{\circ}$&15 $\mu$Jy/beam \\
 & &  &  &  &  & $10 \arcsec \times 10 \arcsec$ &  $0^{\circ}$& 31 $\mu$Jy/beam\\
\hline
\end{tabular}
\label{obs_details}
\end{table*}

\section{PSZ2 G200.95-28.16}
\label{sec:G200}

PSZ2 G200.95-28.16 is a highly disturbed galaxy cluster with a mass $M_{500} = 2.7 \pm 0.2 \times 10^{14}~M_{\odot}$ \citep{https://doi.org/10.1051/0004-6361/201116459} at a redshift $z = 0.22$ \citep{https://doi.org/10.1051/0004-6361/201118731}. It was first discovered with the Planck Satellite \citep{https://doi.org/10.1051/0004-6361/201116459} and later confirmed by XMM-Newton observations \citep{https://doi.org/10.1051/0004-6361/201118731}. The cluster showed a highly disturbed morphology with a flat X-ray surface brightness profile in the XMM-Newton images. The X-ray luminosity in the 0.1-2.4 keV band of the cluster was found to be $(0.99 \pm 0.04) \times 10^{44}\ \mathrm{ergs \ s}^{-1}$ \citep{https://doi.org/10.1051/0004-6361/201118731}. The merging cluster has an offset of 700 kpc between its X-ray and Sunyaev-Zeldovich (SZ) peaks towards the Southwest direction, the highest in the Planck early results catalogue \citep{https://doi.org/10.1051/0004-6361/201118731}.

\citet{2017MNRAS.472..940K} discovered a bright radio relic in the Southwest of the cluster using GMRT (235, 610 MHz) and JVLA (1.5 GHz) observations. At that time, it was the lowest-mass galaxy cluster to host a single radio relic. Merger shocks were not detected in XMM-Newton images. Under the assumption of DSA, the Mach number of the merger shock was deduced to be $3.3\pm 1.8$ from the spectral index of the radio relic. The relic had an arc-like morphology with a notch in the outer edge contributed by a low surface brightness region \citep{2017MNRAS.472..940K}.  \citealp{2017MNRAS.472..940K} argued that the presence of a merging shock can lead to over-pressured regions responsible for the X-ray - SZ offset. Based on the hydrodynamical simulations, investigating the X-ray -SZ offsets in merging clusters\citep{2014ApJ...796..138Z}, the offset was thought to be a product of merging of two comparable mass sub-clusters with a mass ratio
between 1 and 3\citep{2017MNRAS.472..940K}. The discovered radio relic was an order of magnitude more powerful than predicted by the existing correlations, which give a power law dependence of radio power at 1.4 GHz with the cluster mass \citep[$P_{\rm 1.4~GHz} \propto M^{2.83 \pm 0.39}$;][]{https://doi.org/10.1093/mnras/stu1658}. The MeerKAT Galaxy Cluster Legacy Survey(MGCLS; \citealp{2022A&A...657A..56K}) reported two new radio relics in this cluster, one in the East, and another faint small candidate relic in the Northwest. 
In this work, we will present a detailed radio and X-ray band study of the merging cluster.

\section{Radio Observations and Data Reduction:} \label{sec:obs} 

We have observed this cluster with uGMRT at 400, 650, and 1250 MHz respectively on 12, 13, and 14th July 2020 with a total observation time of 8 hours at each frequency. The 1250 MHz data had to be discarded due to the effect of the systematics present at the observatory at that time (uGMRT central square baselines' offset problem\footnote{\url{http://gmrt.ncra.tifr.res.in/gmrt_users/help/csq_baselines.html}, \url{http://www.ncra.tifr.res.in/ncra/gmrt/gmrt-users/recent-gmrt-updates}}
. The baselines involving only the central square antennas had an offset ($\geq 15\%$ at 1250 MHz) in response in the calibrated visibilities compared to the other baselines involving arm antennas. The 400 MHz data were severely affected by radio frequency interference (RFI) and thus we re-observed the cluster at 400 MHz for 10 hours on 23rd July 2023. This observation was also severely affected by RFI and roughly 70\% data were flagged. A summary of the uGMRT observation is provided in Table ~\ref{obs_details}. In addition, we have used the publicly available MGCLS full stokes images at 1283 MHz \citep{2022A&A...657A..56K}.
\par
\begin{figure*}[ht]
\epsscale{1.2}
\plotone{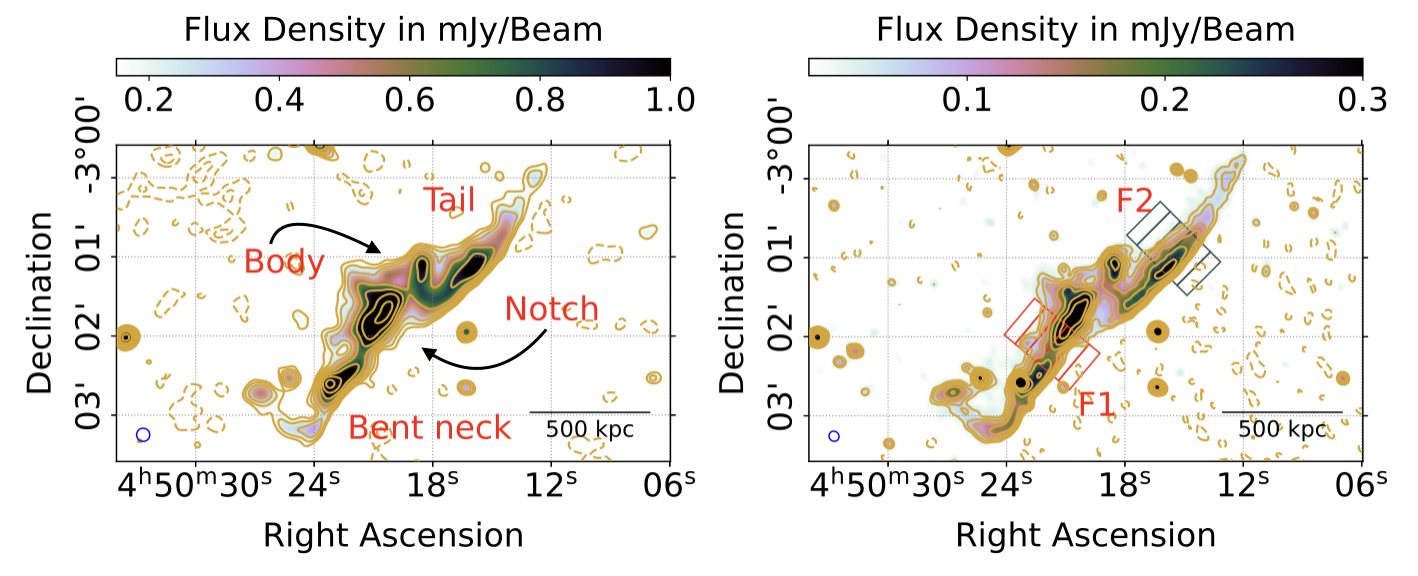}
\caption{
Left: uGMRT $10 \arcsec \times 10 \arcsec$ 650 MHz image of the relic Seahorse. 
Right: $7.9 \arcsec \times 7.3 \arcsec$ MeerKAT 1283 MHz image of the Seahorse. In all cases, the lowest contour starts at 3$\sigma$, and the subsequent levels are plotted in fashion of 3$\sigma \times $(1, $\sqrt{2}$, 2, 2$\sqrt{2}$,...) with $\sigma_{\mathrm{650 MHz}} = 31\mu$Jy/beam, $\sigma_{\mathrm{1283 MHz}} = 8\mu$Jy/beam. The dashed line represents $-3\sigma$ in each case. F1 and F2 represent two filaments in the Seahorse morphology. The red and green boxes represent the region used to derive the surface brightness profiles for the two filaments.}
\label{R1}
\end{figure*}

\begin{figure}
\epsscale{1.1}
\plotone{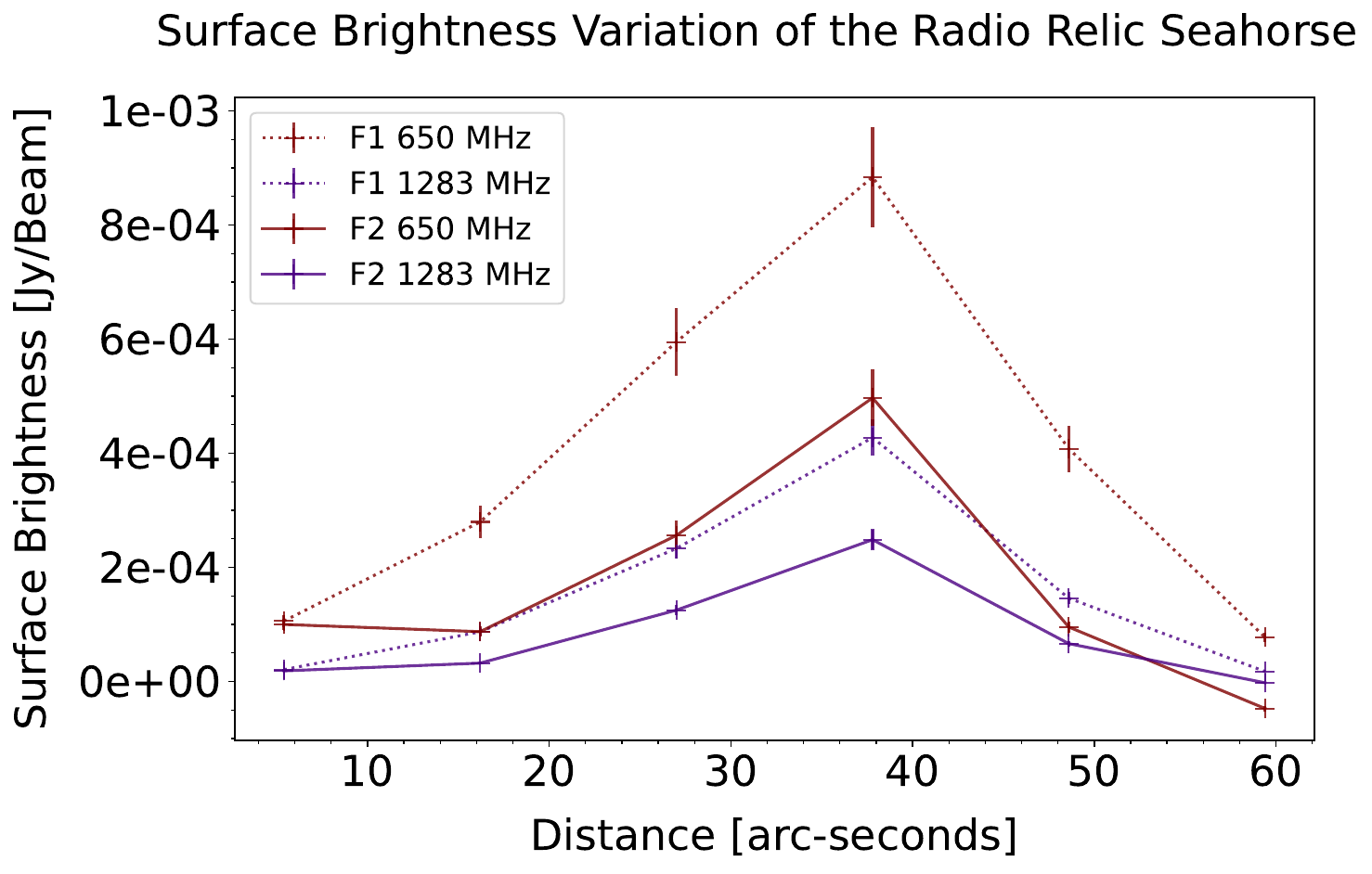}
\caption{Variation of surface brightness across the relic width for the two filaments F1 and F2 using the regions pointed in Fig.~\ref{R1}. The starting point for both of the regions F1 and F2 are in the shock downstream as shown in figure \ref{R1} and each region has a width of $10\arcsec$ and height of $33\arcsec$.}
\label{r1_prof}
\end{figure}

\begin{figure*}
\epsscale{1.2}
\plotone{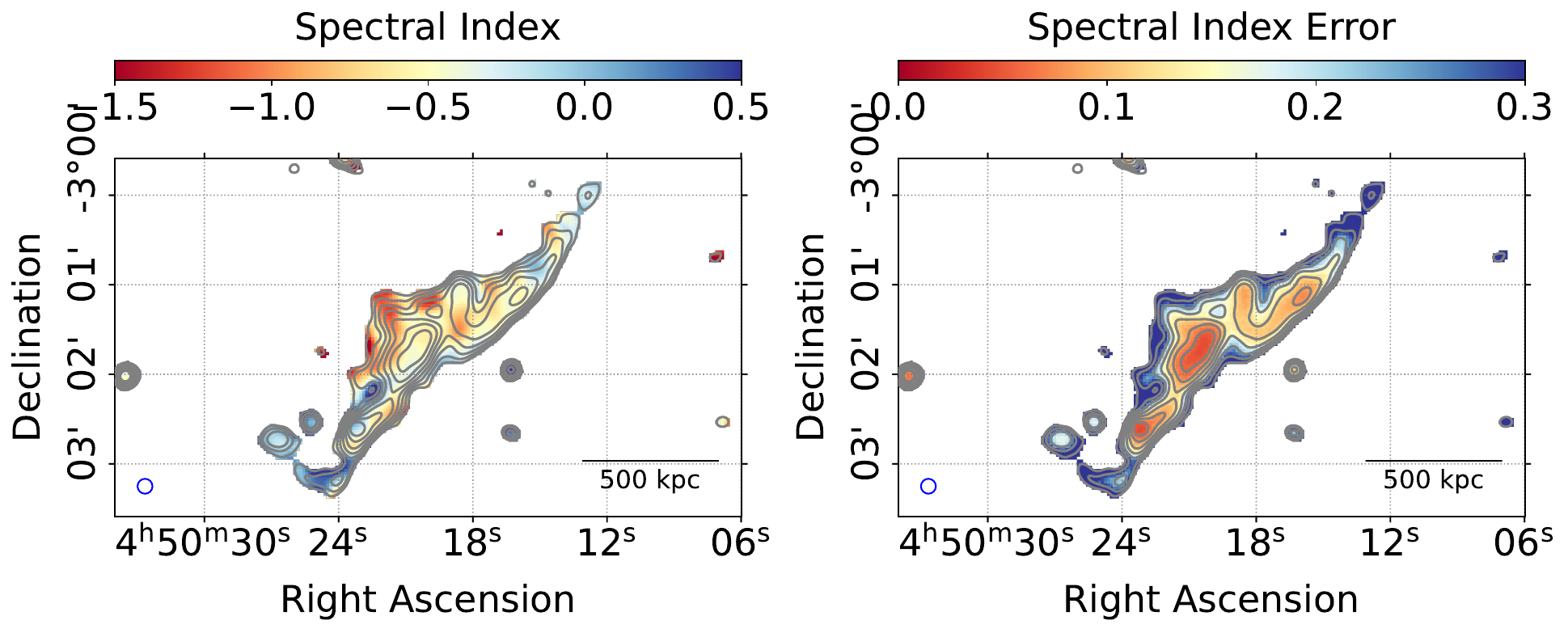}
\caption{Left: Spectral index distribution of the Seahorse radio relic using 650 MHz uGMRT and 1283 MHz MeerKAT images, smoothed to a common resolution of $10\arcsec \times 10\arcsec$. Right: Error in the spectral index distribution maps. The black contours represent contour levels starting at $3\sigma$ with a multiplicative increment of $\sqrt{2}$ in both cases.}
\label{r1spec}
\end{figure*}
\subsection{uGMRT Data Reduction} We have used radio interferometric data analysis software Common Astronomy Software Applications 6.5 (CASA; \citealp{2022PASP..134k4501C}) for data analysis. 3C286 and 3C147 were used for the delay, bandpass, and flux calibration. 3C138 was used as the phase calibrator. To deal with RFI, we have used CASA task flagdata with mode `tfcrop' before calibration and a mix of CASA flagdata modes like `rflag' and `tfcrop' after calibrating the data. We have run the mode `rflag' with baseline-based cut-offs to avoid the biasing of the statistics and to preserve the shorter baselines to avoid the loss of extended emission. After calibration, the source data were split into a separate file and frequency averaged keeping the bandwidth smearing in mind. We have sub-banded the data taking 20 MHz as one sub-band to refine the gain solution in the self-calibration loops. We have used CASA task `tclean' to image the visibilities. Several rounds of phase-only self-calibration followed by a couple of amplitude-phase self-calibrations were carried out on the visibilities to get the final image. After each round of self-calibration, the model was subtracted from the data, and CASA tasks `rflag' and `tfcrop' were applied to this residual data to flag out mid and low-level RFIs. The final calibrated visibilities were imaged using `tclean' with nterms = 2 and a robust value of 0. To study the diffuse emission and look for a faint halo, We first imaged the cluster field by selecting only large baselines greater than 5 k$\lambda$ and selected the point sources inside the cluster field interactively. Then a model was constructed from the point source image and subtracted from the visibilities accounting for the w-terms. The residuals are then imaged with `tclean' with nterms = 2 and robust value of 0. With uGMRT, the average uncertainty of the fluxscale is of the order 10\% \citep{2017ApJ...846..111C}\par

\subsection{uGMRT-MeerKAT Data Post-processing}
We have used the archival MCGLS full Stokes images at 1283 MHz to extend our spectral coverage and explore the polarization properties of the radio relics. For the spectral index maps\footnote{We assume the flux density $\textrm{S}(\nu) \propto \nu^{\alpha}$, where $\nu$ is the frequency and $\alpha$ is the spectral index.}, we have smoothed both uGMRT 650 MHz and 1283 MHz MeerKAT images to a resolution of $10\arcsec \times 10 \arcsec$. Then a 3$\sigma$ cut-off was put on both the images to mask the noise and calculate spectral index distribution for the un-masked regions. The 3$\sigma$ cut-off was also used to define a region around the extended emission and to get the flux densities. We have used the CASA task `imfit' to model the point sources in the field to calculate the flux densities.

In MGCLS, the Q, U maps are made by taking noise-weighted sum from the frequency planes of the Q, U spectral cubes \citep{2022A&A...657A..56K}. The available Q and U images are used to get the fractional polarization maps and distribution of electric vector position angles (EVPA). The fractional polarization is defined as $\frac{\sqrt{Q^2+U^2}}{I}$, while the uncertainty in the fractional polarization is calculated as 
\begin{equation}
    \sigma_p = \frac{1}{I \sqrt{Q^2 + U^2}} \sqrt{Q^2 \sigma_Q^2 + U^2 \sigma_U^2} + \frac{\sqrt{Q^2 + U^2}}{I^2} \sigma_I
\end{equation}. 
Here, $\sigma_I, \sigma_Q,\text{ and } \sigma_U$ represent the errors on the flux densities of Q, U and I and are calculated as a $\Delta F = \sqrt{(0.1F)^2+(N_{beams}\times r.m.s.)^2}$. F is the flux density of the interested region, $N_{beams}$ is the number of beams covered in the region and r.m.s. is the sensitivity of the relevant image. Using the spectral cube (12 frequency planes from 918 to 1656 MHz) from MGCLS \citep{2022A&A...657A..56K}, we run RM synthesis \citep{2005A&A...441.1217B} using the RM-tools package\citep{ 2020ascl.soft05003P} to get the RM and intrinsic fractional polarization over the radio relic surface. The galactic rotation measure (RM) contribution in the cluster direction\footnote{We used the CIRADA RM cutout server at \url{http://cutouts.cirada.ca/rmcutout/} \citep{https://browse.arxiv.org/pdf/2102.01709.pdf}} is $4.2 \pm 9.4 \text{ (dispersion within the cluster region)~rad~m}^{-2}$. We corrected for the average galactic RM contribution along the line of sight using the average value.
\section{X-ray Data Analysis}
\label{sec:X-ray}

The Chandra Advanced CCD Imaging Spectrometer (ACIS) observations used in this study (ObsID 19566, 20852, 20853, 20854, 20855, and 20858) were conducted between November 19 and 24, 2017, totalling 152 ks of exposure time. The data processing was performed using CIAO v4.11 (2018 Dec 5, \citealp{2006SPIE.6270E..1VF}) and CALDB v4.8.2. Standard event filtering procedures were applied to mask bad pixels by removing cosmic ray afterglows and streak events and eliminating detector background events identified using the VFAINT mode data. The data were further analyzed to check for background flares. This was done using the 2.5–7 keV light curve in 1 ks time bins in a region free of cluster emission and excluding point sources. Additionally, the ratio of 2.5–7 keV to 9.5–12 keV counts was used as a more sensitive check for faint flares. The background count rates were found to be steady without any strong flares, which are typically characterized by a $\geq 20\%$ deviation from non-flaring time bins or a sustained increasing or decreasing trend. Therefore, no additional time periods were excluded from the analysis.

We identified point sources visually from soft and hard band images, which were binned and smoothed at several different scales; we remove them from the images considered here.
Blank sky data sets from the Chandra Calibration Database were identified and matched to our observations following the methods outlined in \citealp{2003ApJ...583...70M} and \citealp{2006ApJ...645...95H}. The ACIS background period relevant to these observations is period G, with a total exposure of 600 ks for the ACIS I0, I1, I2, and I3 CCDs. We ignore data from the CCDs in the ACIS-S strip, as they are located more than 8\arcmin\ from cluster center and contain no appreciable signal. The background was scaled to these observations by the ratio of 9.5--12~keV counts across the field of view, which accounts for variations in the background rate over time. This method provides an accurate reconstruction of the background level across the 0.8--9~keV energy band to 3\%, at the 90\% confidence interval \citep{2006ApJ...645...95H}.
We model the ACIS readout artifact by creating an image from the events file after randomizing their chip coordinates along the readout direction.
This image is then scaled by the ratio of the readout time to the sum of the readout time and the frame time, and subtracted from the original. 
This procedure is
consistent with the approach found in \citep[e.g.,][]{2000ApJ...541..542M}.

A companion paper presenting a more complete analysis of the X-ray data is currently in preparation.
This follow up work will present a search for a shock front across the primary relic and further investigate the temperature structure of the hot phase of the ICM.
\begin{figure}[ht!]
\epsscale{1.2}
\plotone{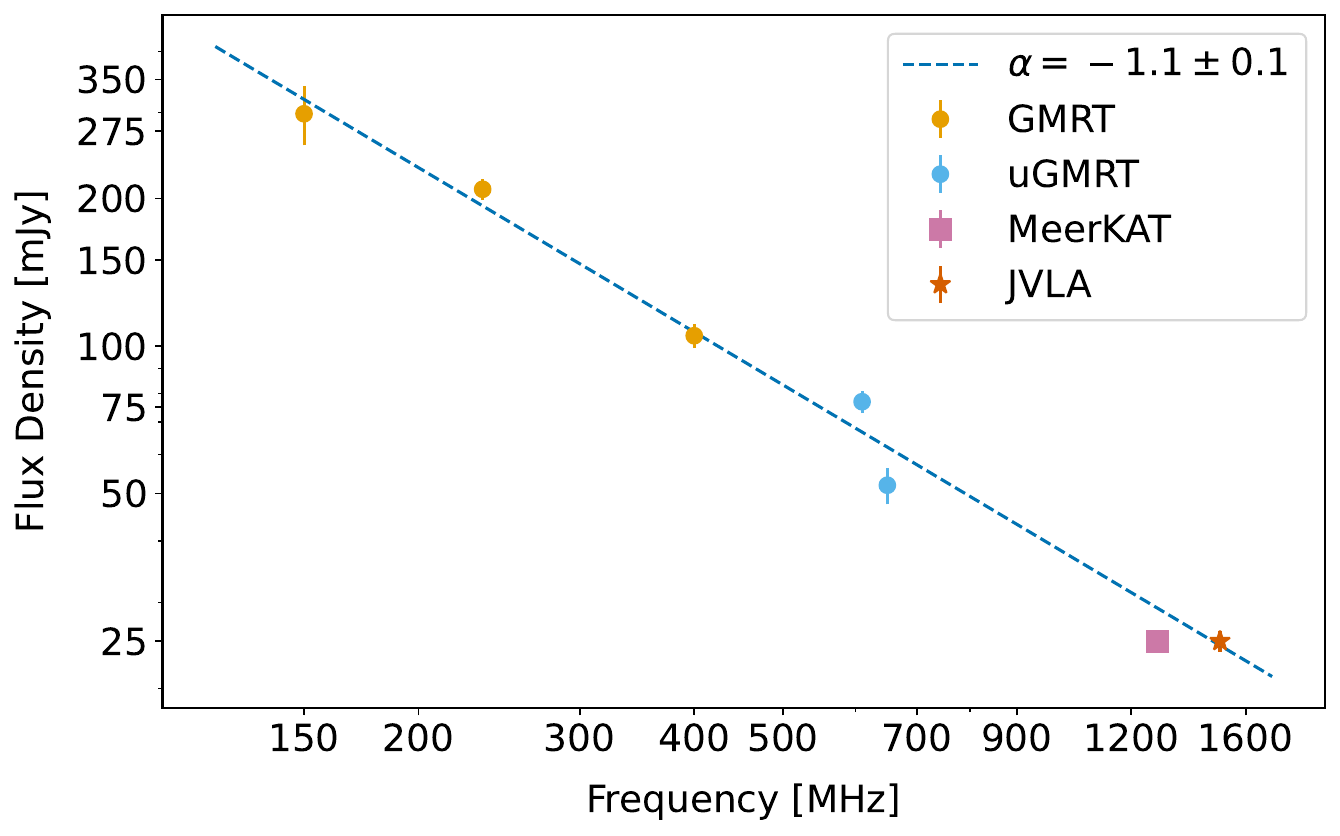}
\caption{Integrated spectral index for Seahorse radio relic using all the available data points from \citealp{2017MNRAS.472..940K} and new measurements in this work.}
\label{int_spec}
\end{figure}

\section{Results} \label{sec:res}
\subsection{Radio and X-ray Images} 
The uGMRT image along with the X-ray and optical images are shown in Fig.~\ref{rgb}. The cluster is highly dynamic and consists of two subclusters. The uGMRT 400 and 650 MHz images are presented in Fig.~\ref{field}. The brightest radio relic can be seen as an elongated structure in the Southwest direction (R1). The bent and elongated morphology is seen in both high and low-resolution images. The morphology of R1 resembles a seahorse and will be referred to as `Seahorse' throughout the text. The high-resolution image at 650 MHz (Fig.~\ref{field} top right) hints at another diffuse radio emission to the East of the cluster which is the second radio relic in the system, R2. In the low-resolution images at uGMRT 650 MHz (Fig.~\ref{field}, bottom left) and MeerKAT 1283 MHz (Fig.~\ref{field}, bottom right), another small diffuse radio emission, R3 can be traced out in the Northwest direction.
Both R2 and R3 are recovered in the low-resolution 650 MHz images (Fig.~\ref{field} bottom left). The X-ray surface brightness maps overlayed with MeerKAT 1283 MHz contours in Fig.~\ref{field} (Bottom right) show all three radio relics. 
\par
The relevant point sources in the cluster field are labelled with S1-S5 and A, B on top of the uGMRT 650 MHz image (Fig.~\ref{field}, top right). Flux densities of all the mentioned point sources are listed in table ~\ref{point}. The properties of the three relics are summarised in Tab.~\ref{relic_prop}.

\begin{table}
\centering
\caption{Flux densities and spectral indices of the discrete sources labelled in Fig.~\ref{field}.}
\begin{tabular}{cccccc}
\hline
\hline
  Label  & $\mathrm{S}_{650\mathrm{MHz}}$ (mJy) &  $\mathrm{S}_{1283\mathrm{MHz}}$ (mJy)                & $\alpha$              \\
\hline
S1 & $89 \pm 9$  & $49 \pm 5$ & $-0.8\pm 0.3$\\
S2 &  $38 \pm 4$& $18 \pm 2$ & $-1.1\pm 0.4$\\
S3 & $14 \pm 1 $& 8 $ \pm 0.8 $& $-0.9 \pm 0.4$ \\
S4 & $3 \pm 0.3$& 1.8$\pm 0.2$& $-0.7 \pm 0.3$ \\
S5 & $3.8 \pm 0.4 $& $1.2 \pm 0.1$&$-1.7\pm 0.6$\\
A & $1.6 \pm 0.2$ & $1.1 \pm 0.1$&$-0.6\pm 0.2$ \\
B & $0.9 \pm 0.1 $ & $1.2 \pm 0.1 $ &$+0.5\pm 0.2$\\
\hline
\end{tabular}

\label{point}
\end{table}

\subsection{Radio Relic Seahorse}
The Seahorse is the brightest among all three radio relics in the system. We confirm the presence of the notch that was already reported by \citealp{2017MNRAS.472..940K}.
The relic morphology consists of two bright filaments separated by the notch. The relic has a bent head to the east, a notch in the body and an elongated tail (Fig.~\ref{R1}, left). 
The surface brightness profiles across the width of the radio relics  (Fig.~\ref{R1}, right) are typically well-modelled by a log-normal distribution. The surface brightness gradually increases up to the location of the shock front and then shows a steeper fall (Fig. \ref{r1_prof}). In both MeerKAT 1283 MHz and uGMRT 650 MHz images, the two filaments of Seahorse have similar `typical' radio relic-like surface brightness profiles, with coinciding locations of the peaks (Fig.~\ref{r1_prof}). At the same resolution ($10\arcsec \times 10\arcsec$) images, the uGMRT 650 MHz surface brightness profiles are wider compared to the MeerKAT 1283 MHz images , indicative of larger shock downstream region in the lower frequencies. The radio relic emission due to its synchrotron origin is expected to have a larger width in the lower frequencies. With the resolved spectral index maps (Fig.~\ref{r1spec}), the Seahorse radio relic shows spectral steepening towards the cluster centre and is indicative of the direction of merger shocks. The spectral index of the bent head of the Seahorse is flatter compared to the rest of the emission. As there are no optical counterparts near the neck, the possibility of mixing radio galaxy lobes with the relic emission at the bent head is unlikely. The Seahorse has an integrated spectral index of $-1.1\pm 0.1$ using all the available data from \citep{2017MNRAS.472..940K} and new observations (Fig.~\ref{int_spec}). 
Point source `A' (Fig.~\ref{field}, top right) visually overlaps with the relic emission and is a galaxy named 2MASX J04502353-0302367 with a K band magnitude of 12.94 \citep{2006AJ....131.1163S}. Using the K band luminosity and redshift correlation \citep{2003MNRAS.339..173W, 2004MNRAS.348..165E}, that translates to a redshift of 0.12, lower as compared to our cluster redshift of 0.22.

\subsection{Radio Relics R2 and R3}
\begin{figure*}[ht!]
\epsscale{1.2}
\plotone{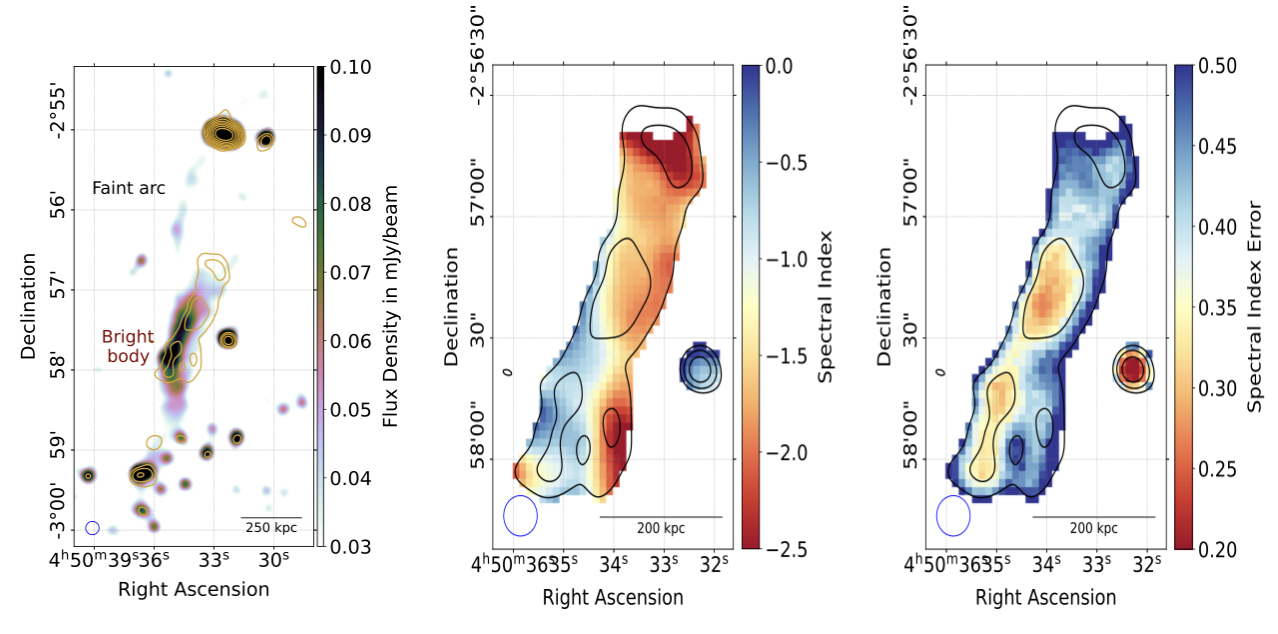}
\caption{Left: uGMRT 650 MHz $10 \arcsec \times 10\arcsec $ contours in yellow overlayed on MeerKAT smoothed 1283 MHz $10 \arcsec \times 10\arcsec$ image of relic R2 in colour scale. The lowest contour starts at 3$\sigma$, and the subsequent levels are plotted in fashion of 3$\sigma \times$(1, $\sqrt{2}$, 2, 2$\sqrt{2}$,...) with $\sigma_{\mathrm{650 MHz}} = 45\mu$Jy/beam. Middle: Spectral index distribution of the relic R2 using 650 MHz uGMRT and 1283 MHz MeerKAT images, both smoothed to a common resolution of $10\arcsec \times 10\arcsec$. Right: Error in the spectral index distribution maps. The black contours represent contour levels starting at $3\sigma$ with a multiplicative increment of $\sqrt{2}$ in both cases. }
\label{r2spec}
\end{figure*}

In MeerKAT 1283 MHz images, R2 is seen to have two substructures, a wider bright diffuse structure and a faint $\sim 2\sigma$ filament connecting the wider diffuse structure and the point source in the north (Fig.~\ref{r2spec}  and Fig. ~\ref{field} top right). With our uGMRT 650 MHz images, the faint extension is not recovered but the brighter portion is recovered and two filamentary tails can be traced towards the south. The emission in the brighter portion of the R2 is wider at 650 MHz compared to 1283 MHz. For the brighter portion, the integrated spectral index of the radio relic is $-1.4 \pm 0.2$. In the resolved spectral index maps, the radio relic R2 is seen to exhibit spectral steepening towards the cluster centre and also across the relic length towards the north (Fig.~\ref{r2spec}). 

The radio relic R3 is the smallest radio relic (Tab.~\ref{relic_prop}) in the system, situated to the north of the cluster. The relic R3 is arc-like. In our 650 MHz smoothed ($10 \arcsec \times 10 \arcsec$) image (Fig.~\ref{field}, bottom left), fractions of R3 with an extension towards the west are found. The integrated spectral index of the R3 is $-1.6 \pm 0.3$.\par

\subsection{Polarization measurements} 
\begin{figure*}[ht!]
\epsscale{1.2}
\plotone{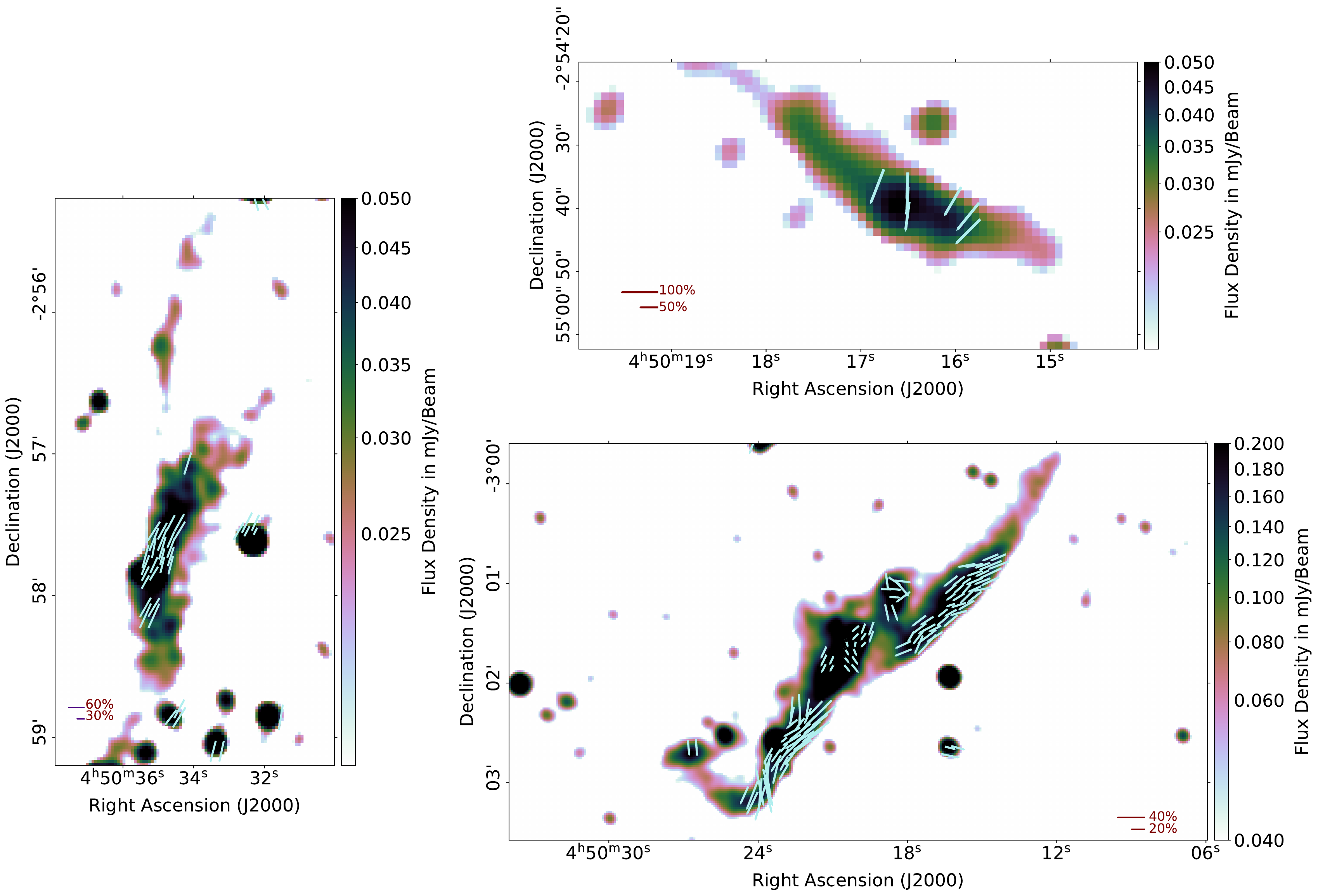}
\caption{MeerKAT 1283 MHz polarization vector (B mode) scaled with fractional polarization plotted over the intensity distribution for the relics Seahorse, R2, and R3 in colour scale. The B vectors are corrected for galactic RM contributions. Regions having values more than $3\sigma$ in total intensity and $5\sigma$ in linear polarization intensity are only taken for consideration while plotting the vectors.} 
\label{rpol}
\end{figure*}

\begin{figure*}[ht!] 
\epsscale{1.2}
\plotone{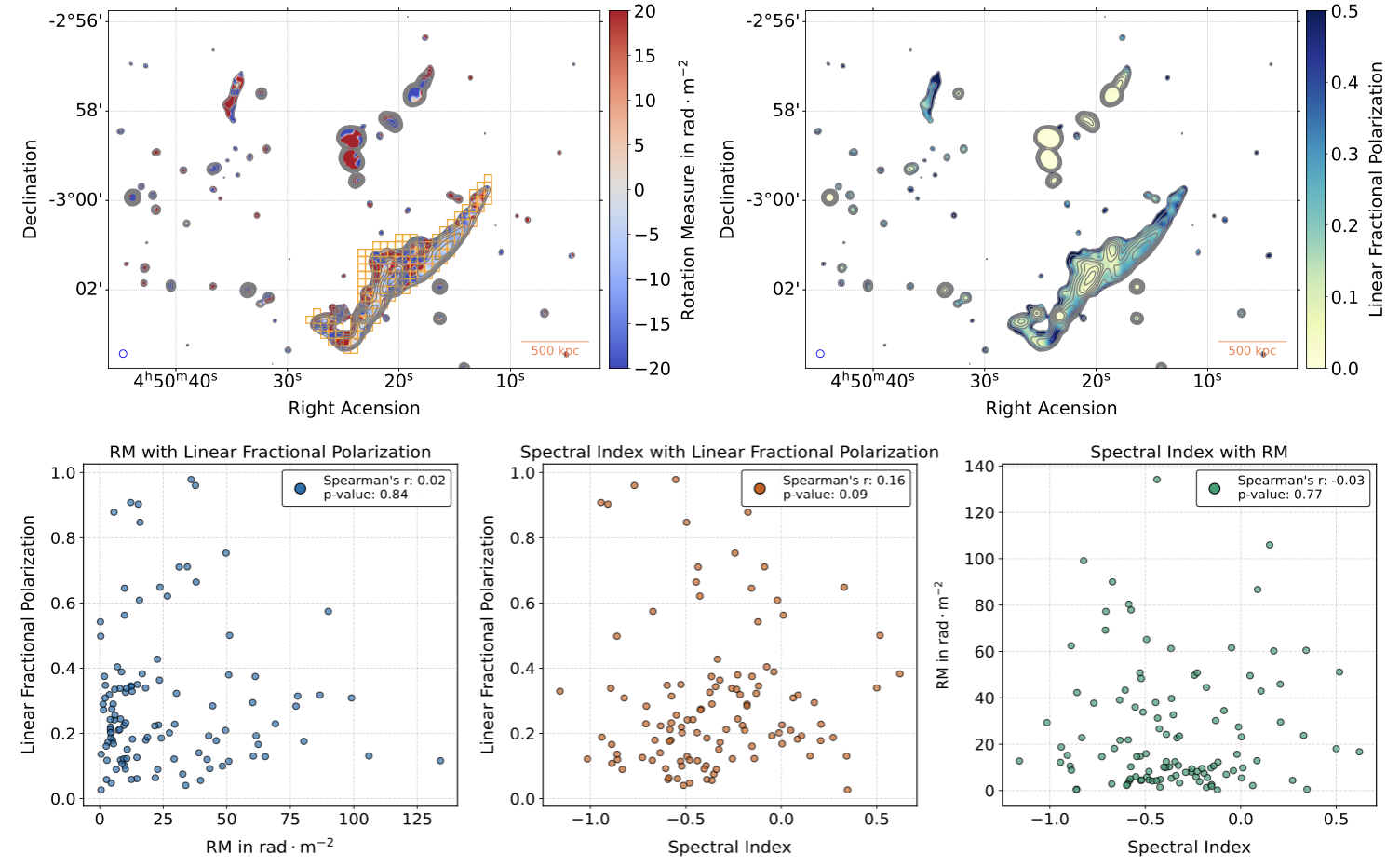}
\caption{Top Left: 10$\arcsec \times 10\arcsec$ RM map of the PSZ2 G200.95-28.16 system in colour scale. The coral squares with 10$\arcsec$ width and height represent regions over which the RM, fractional polarization, and spectral index maps are averaged for the correlation analysis in the bottom panels. Top Right: Intrinsic fractional polarization maps of the seahorse radio relic at 10$\arcsec \times 10\arcsec$ resolution. Bottom Panel: Displays the correlation analysis between RM and linear fractional polarization, spectral index and linear fractional polarization, and spectral index and RM. The Spearman's correlation coefficient (r) and the associated p-value are indicated in the legend.}
\label{RM}
\end{figure*}

We have used the $\sim 8\arcsec$ resolution MGCLS \citep{2022A&A...657A..56K} Stokes Q, U images to deduce the polarization information. The fractional polarization maps are shown in Fig.~\ref{rpol}. All of the three relics are linearly polarized, and the relic R2 and R3 have comparatively stronger linear polarization than the Seahorse relic. We have shown the polarization B vector distribution in figure \ref{ring} after correcting for the average galactic RM and length of the ticks scaled with fractional polarization. The B vectors in the Seahorse radio relic exhibit a smooth and aligned distribution following the radio surface brightness in both the filaments `F1' and `F2'. The Seahorse shows a depolarization across the shock width. The radio relic R2 also has an aligned magnetic field distribution while for R3 the B vectors are somewhat normal to the shock surface but given the low surface brightness of the radio relic R3, the structures of B distribution in the radio relic remain inconclusive. The average fractional polarization of the Seahorse radio relic is $23 \pm 2\%$ while the average fractional polarization of relics R2 and R3 are $28 \pm 4\%$ and $58 \pm 9\%$. The RM maps are patchy, with a mean RM of -1.25 $\text{rad} \cdot \text{m}^{-2}$ and a dispersion of 16.47 $\text{rad} \cdot \text{m}^{-2}$. We have smoothed the RM and fractional polarization maps to a common resolution of $10\arcsec\times10\arcsec$ and then distributed square regions of $10\arcsec$ resolution over the whole radio relic emission (Fig \ref{RM})to study the correlations between intrinsic fractional polarization, spectral index and RM. We have found a very weak correlation between the spectral indices and linear fractional polarization with a linear slope of 0.16 and p-value of 0.09, correlating the depolarization and the spectral steepening at the downstream region. The complex morphology of the radio relic emission is probably also responsible for the non-significance fo the correlation.

\subsection{A Radio Ring} 
\begin{figure*}[ht!]
\epsscale{1.2}
\plotone{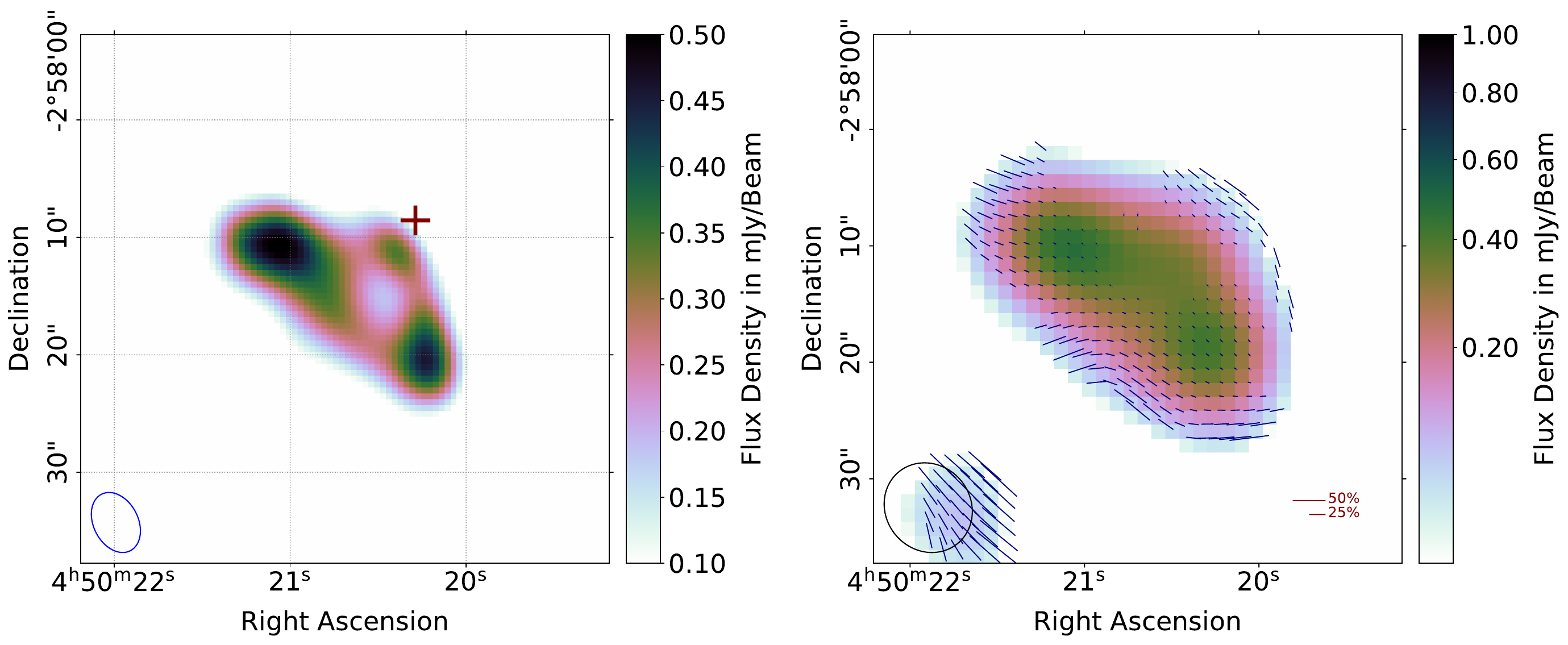}
\caption{Left: uGMRT 650 MHz $5.4 \arcsec \times 3.8 \arcsec$ image of the radio ring(`S4'). The `+' shows the position of the very faint galaxy associated with the ring. Right: The polarization vector (B mode) scaled with fractional polarization plotted over the intensity distribution. The red lines at the bottom signify the tick lengths at 25 \% and 50\% fractional polarization.}
\label{ring}
\end{figure*}

In the high-resolution uGMRT images, we have also detected a ring-like radio source. The source is labelled as `S4' in Fig.~\ref{field}. The flux densities of the ring are 2.92 mJy and 1.76 mJy at 650 and 1283 MHz. The ring approximately has a length of 100 kpc with a width of 65 kpc. The resolution is not sufficient to study the spatially resolved spectral properties but the mean spectral index of the radio ring is $-0.7 \pm 0.1$. For the radio ring 'S4' the B vectors and fractional polarization profile exhibit smooth distribution with edges being more linearly polarized than the core (figure \ref{ring}).


\begin{table*}
\centering
\caption{Properties of the radio relics. Notes: While calculating the largest linear size (LSS) for radio relic R2, we have included the faint extension of R2 in the MeerKAT image which is not detected in uGMRT band 4. To calculate the flux densities, we have taken $3\sigma$ cut-off on both frequencies and calculated the flux densities. We have reported the point source subtracted flux densities.
}
\begin{tabular}{lccc}
\hline
\hline
    & Seahorse               & R2                 & R3                    \\
\hline
LLS & 1.53 Mpc         & 1.12 Mpc           & 340 kpc               \\
$S_{650 \ MHz}$   & $51 \pm 9$ mJy      & $6.95 \pm 0.90$ mJy    & $1.64 \pm 0.30$ mJy         \\
$S_{1283 \ MHz}$   & $25 \pm 2.3$ mJy    & $2.70 \pm 0.20$ mJy    & $0.55  \pm 0.10$ mJy         \\
$P_{1.4 \ GHz}$  & $(4.3 \pm 0.5)\times 10^{24} $W Hz$^{-1}$ & $(5.8 \pm 0.7)\times 10^{23}$W Hz$^{-1}$ & $(9.92\pm1.1)\times10^{22}$W Hz$^{-1}$\\
Largest width   &  237 kpc  & 131 kpc  & 67 kpc\\   
$\alpha_{1283 \ MHz}^{650 \ MHz}$   &  $-1.1\pm 0.1 $  & $-1.4 \pm 0.2 $  & $-1.6 \pm 0.4$\\
\hline
\end{tabular}

\label{relic_prop}
\end{table*}
 
\section{Discussion}\label{sec:discussion}
\subsection{Cluster Mass and Radio Power}
\begin{figure*}[ht!] 
\epsscale{1.2}
\plotone{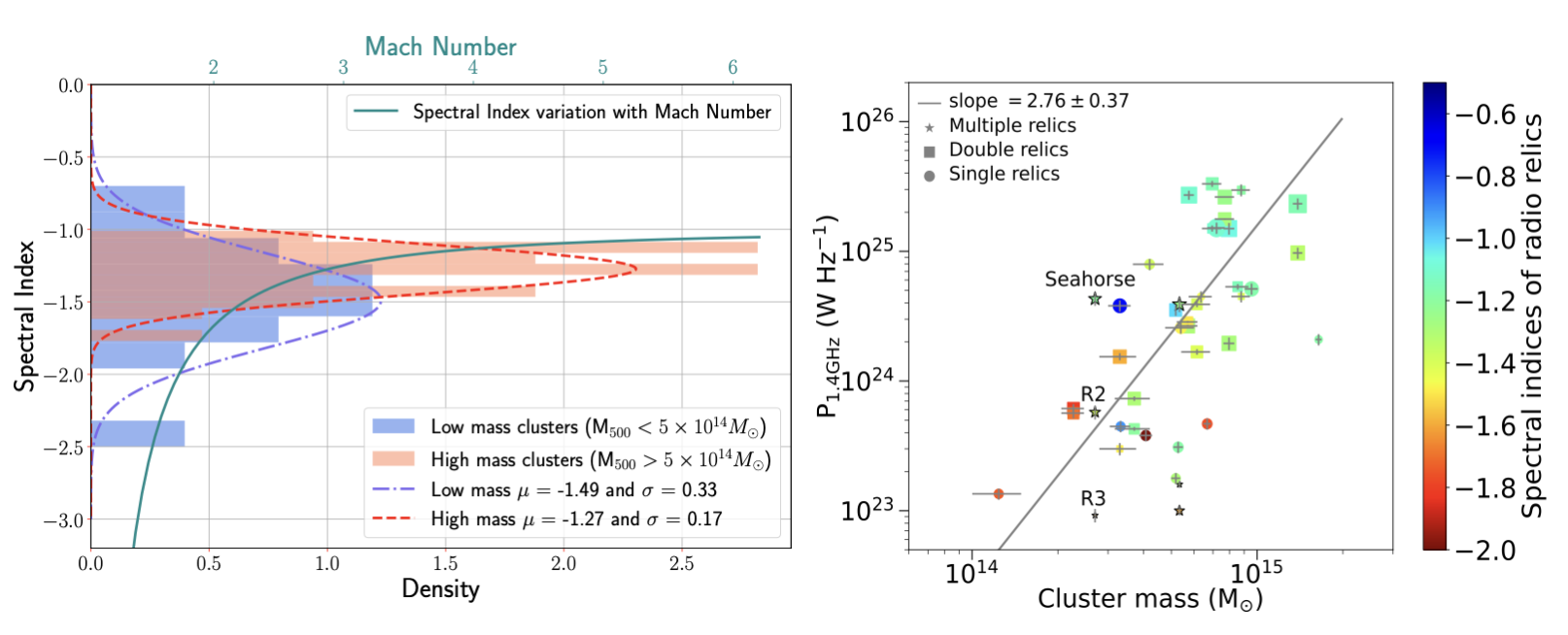}
\caption{
Left: The red histogram shows the distribution of integrated spectral indices in the high-mass cluster sample while the blue represents the low-mass galaxy cluster sample. The dashed red and green lines represent the fitted Gaussians to the high-mass and low-mass spectral index distributions, respectively. The solid green line represents the expected spectral index at a given Mach Number under the assumption of DSA with continuous injection of electrons. We have chosen the Mach Number from 1.6 to 6 to stay consistent with the observed values in radio relics and to match the spectral indices observed in the sample. Right: Cluster mass and Radio power at 1.4 GHz plotted for different single and double relics with available spectral indices. The line represents a correlation slope of 2.76. The spectral indices are shown in colour scale. The three radio relics are marked with Seahorse, `R2' and `R3'. Double relics are shown in squares, single relics are in circles and multiple relic systems, Abell 746 and PSZ2 G200.95-28.16 are shown in star markers. The data points are taken from \citealp{ https://doi.org/10.1093/mnras/stu1658,2018MNRAS.477..957D}.}
\label{mass-power}
\end{figure*}

A scaling relation between radio power of radio relics and host cluster mass can provide insights into the energetics of the merger.
\citealp{ https://doi.org/10.1093/mnras/stu1658} found a power law scaling relation with a slope of $\sim 2.8$ between the radio luminosity of the relic at 1.4 GHz and the mass of the host cluster. With a new sample from LOFAR LOTSS DR2, using both single and double relics, \citet{https://arxiv.org/pdf/2301.07814.pdf} obtained a steeper slope of around 5.5. The simulations with the assumption that a fixed fraction of merger kinetic energy is dissipated through the shocks which are responsible for the relic emission \citep{2006MNRAS.373..881P}, hint at a much flatter correlation slope of 5/3 \citep{ https://doi.org/10.1093/mnras/stu1658}. So far the multiple relic systems have not been included in the scaling relation. 
We plotted the three relics in PSZ2 G200.95-28.16 on the radio power- mass plane (Fig.~\ref{mass-power}). While R2 is closest to the scaling relation, Seahorse and R3 contribute to the scatter of the correlation. The low-mass system has a variety of radio relics in terms of the radio powers observed.
\par

We have plotted another multiple radio relic system Abell 746 \citep{2023arXiv230901716R} on the radio power at 1.4 GHz--cluster mass plot (Fig.~\ref{mass-power}).  
While the NW relic seems to abide by the correlation, the other two radio relics are under-luminous given the mass of the merging cluster. Based on the merger parameters, the produced shocks in a single system can have different strengths to (re)accelerate the CRes. The difference in the strengths of the produced shocks in complex systems can be attributed to the scatter of the correlation between mass and radio power at 1.4 GHz. As in simulations \citep{2024A&A...686A..55L}, though the brightest radio relic power increases with mass, the large abundances of the faint radio relics in every mass range, which are possible candidates to be detected in future radio surveys, challenge the existing correlation, limiting to only the brightest radio relics in the system.\par

The radio synchrotron emission responsible for the relics is generated by the acceleration of CRes in the magnetic fields. As suggested by  \citealp{https://iopscience.iop.org/article/10.3847/1538-4357/ab3983}, following the first core passage of a merger, shear flows occur along the cold fronts and ram-pressure stripped gas. This interaction amplifies and extends magnetic field lines, creating long, smooth magnetic structures spanning $\sim 1-2$ Mpc. These structures are more easily sustained in off-axis mergers. In a low-mass cluster merger, if the subclusters have a sufficient mass ratio, the shocks produced before the low-mass subcluster can have sufficient energy to produce powerful radio relics. The effects of merger parameters like mass ratio, impact parameter, and configuration of magnetic fields can produce merger shocks with a variety of strengths to (re)accelerate the CRes which can manifest themselves as both over and under-luminous radio relics in the radio power-mass correlation. This discussion leaves space for exploring the merger dynamics with different mass ratios and impact parameters in low-mass merger simulations and accounting for redshift dependence in the mass sample.
 \par
We have examined the spectral indices of the relics plotted in the relic power - mass plane (Fig.~\ref{mass-power}). 
Visual inspection shows that the relics in higher mass clusters have spectral indices confined to a narrow range as compared to the scatter seen in the relics in the lower mass clusters. We divided the sample into low and high mass bins separate at $5 \times 10^{14}$~M$_{\odot}$.
The low-mass spectral index distribution has a Gaussian mean of $-1.42$ and a standard deviation of 0.31 while the high-mass end of the spectral index distribution has a mean of $-1.25$ and a standard deviation of 0.17 (Fig.~\ref{mass-power}). Under the assumption of the DSA, the injection spectral index $\alpha_{\mathrm{inj}}$ is related to the Mach number by the equation \citep{1987PhR...154....1B}
\begin{equation}
    \alpha_{\mathrm{inj}} = \frac{M^2+3}{2(1-M^2)}
\end{equation}
In the case of continuous injection at the shock, the integrated spectral index is expected to follow $\alpha_{\rm int} = \alpha_{\mathrm{inj}} - 0.5$. As the Mach number increases, the integrated spectral index initially rises rapidly for $\mathrm{M} \leq 3$, and then continues to increase at a slower rate (green solid line, left panel in Fig.~\ref{mass-power}). 
Higher-mass cluster mergers are expected to generate higher Mach number shocks ($\mathrm{M} \geq 3$) compared to those in low-mass clusters. These stronger shocks are more efficient at accelerating particles, leading to a flatter distribution of spectral indices in radio relics. In contrast, the weaker shocks produced in low-mass clusters can explain the relatively steeper spectral indices observed in their radio relics compared to those in higher-mass mergers. The larger spread in integrated spectral indices can be attributed to the higher rate of change of the integrated spectral index (green line in left panel of Fig. \ref{mass-power}) in the low Mach number (M$\leq3$) regime. In this regime, even small changes in Mach number can significantly alter the integrated spectral index, whereas for shocks with $\mathrm{M} \geq 3$, the index remains relatively stable. Additionally, the different sample sizes—14 relics in the low-mass regime and 23 in the high-mass regime—contribute to the observed scatter in the distributions. This behaviour highlights the differences in spectral indices between relics in low- and high-mass clusters. However, we note that the spread is also influenced by other factors, such as the age of the relic, the duration for which the shock powers the relics, and the redshift of the cluster. Expanding the sample of relics with accurate spectral index measurements and comparing them with MHD simulations will provide deeper insight into the spectral index distributions of radio relics in merging clusters.
\subsection{The Shocks in PSZ2 G200.95-28.16}
The radio relics are typically associated with weak shocks with radio-derived Mach numbers in the range of 2.5 - 4. \citep{2011A&A...528A..38V,2017MNRAS.471.4747C,2017A&A...600A..18K,2017MNRAS.467..936G,2018ApJ...865...24D,2019A&A...622A..21H,2020A&A...642L..13R,2020A&A...634A..64B}. 
For the Seahorse radio relic, the injection spectral index is $-0.7 \pm 0.1$. Under the assumption of DSA, it implies a Mach number of $3.1 \pm 0.8$. The radio relic R2 has an injection spectral index of $-0.8 \pm 0.2$ which corresponds to a Mach number of $2.8 \pm 0.9$. 
\begin{figure*}[ht!] 
\epsscale{1.2}
\plotone{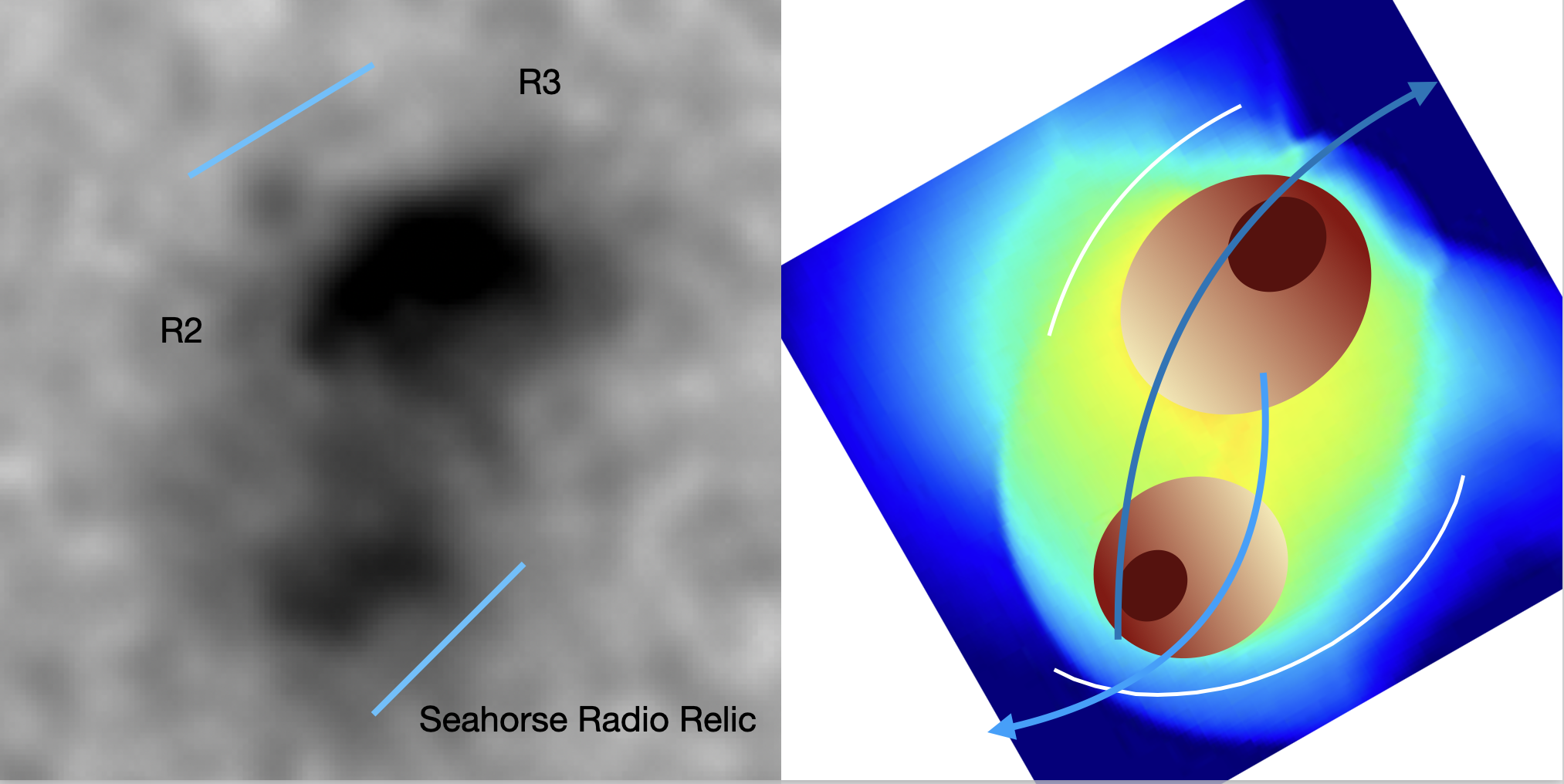}
\caption{Left: X-ray surface brightness image of the cluster PSZ2 G200.95-28.16 with blue lines depicting possible shock locations. Right: A schematic cartoon diagram showing the possible off-axis merger geometry in the system. White curves trace out the approx position of the radio relics and blue arrows depict the motion of the sub-clusters in the merging systems.
}
\label{morphology}
\end{figure*}
\subsection{Merger Geometry in PSZ2 G200.95-28.16}
The morphology of X-ray surface brightness and of the radio relics can be used to propose the merger geometry.  The Seahorse radio relic consists of two bright filamentary sub-structures. 
 The B vector distribution in the MeerKAT full Stokes images (Fig.~\ref{rpol}) shows two distinct distributions in the two filaments `F1' and `F2' within the Seahorse radio relic following the radio surface brightness. Both filamentary shock morphology and magnetic field structures can give rise to filamentary morphology in relics \citep{2023MNRAS.523..701W}, as seen in the Seahorse. The notch and the surface brightness profile of the relics hint towards a merger shock propagation with complex shock morphology.
\par
When two galaxy clusters merge, the produced shocks can be classified as axial and equatorial shocks based on their propagation directions. At the early stages of the merger, the equatorial shocks are generated which expand outwards in the equatorial plane \citep{2018ApJ...857...26H}. Later in the merger stages, the axial shocks are formed which propagate in the merging direction \citep{2018ApJ...857...26H}. The X-ray surface brightness is elongated in the North-South direction connecting the two sub-clusters, hinting a merger along that direction. The axial merger shocks are expected to be oriented perpendicular to the elongation axis. Though no shock is detected in the North-South direction, the non-detection can be attributed to low sensitivity or a small projection angle which is responsible for smearing out the shocks. In our case, the radio relics are not situated perpendicular to the X-ray elongation axis. However, the radio relics Seahorse and R2 are seen to exhibit spectral steepening towards the cluster centre with aligned B vectors over the relic emission, which signifies the propagation of shocks in the outward direction from the cluster. We will discuss some of the possible merging scenarios based on the observed properties of the radio relics.
\begin{enumerate}[label=(\roman*)]
\item Are axial shocks responsible for the radio relics? 

The two of the bright relics, the Seahorse and R2, are close to the equatorial plane if the X-ray elongation axis is assumed to be the merger axis. In simulations, \citealp{2018ApJ...857...26H} showed that the equatorial shocks had higher speeds and Mach numbers compared to axial shocks as they travel through the low-density equatorial regions. But the kinetic energy and cosmic ray energy flux density averaged over the shock surface, for the equatorial shocks are lower. Axial shock in front of the lower mass clump is responsible for most of the energy dissipation and production of CRs. These axial shocks have the best chance to get observed as radio relics and surface discontinuities in X-rays \citep{2018ApJ...857...26H}. The fact that the entire emission of Seahorse and R2 is accounted for by the equatorial shocks is most unlikely given their positions compared to the X-rays. There is a possibility that the Northern tailed portion of the Seahorse and the Southern emission of R2 could result from equatorial shocks. Seeing both axial and equatorial shocks manifesting as radio relics is highly improbable given the merger and synchrotron time scales \citep{2018ApJ...857...26H}. 
\item Is it an off-axis merger? 

In off-axis mergers, the merging sub-clusters merge with a non-zero impact parameter. In X-rays, we observe the presence of two sub-clusters in cases like this, displaying a subtle rotational motion to the centre of mass (Fig \ref{morphology}). In PSZ2 G200.95-28.16, the X-ray morphology reveals two distinct sub-clusters with a potential rotational interaction between them (Fig.~\ref{morphology}).
 The morphology is similar to the simulated off-axis cluster merger in \citealp{2020ApJ...894...60L} (surface brightness maps in Fig. 2, at t = 0.4 Gyr). The smaller sub-cluster is expected to produce the most luminous radio relic. As the pericentre distance is higher, the cool-cores of the clusters were not heated up due to the merger and can be traced out in simulated cluster merger maps.\par
Due to the whirling motion of sub-clusters in off-axis mergers, the axial merger shocks are traced with an offset to the elongation axis which can explain the position of radio relics in PSZ2 G200.95-28.16. The faint extension ($2\sigma$) of R2 towards the north 
and then a broken arc up to R3 (Fig.~\ref{field}, bottom right) can be produced by the weaker shock front produced by the larger sub-cluster while the Seahorse radio relic is the result of the stronger shock produced in front of the smaller sub-cluster (Fig \ref{morphology}). The notch and the existence of filaments in the Seahorse radio relic can be explained by the complex structures of the produced shock surface or the distribution of magnetic fields \citep{2023MNRAS.523..701W}.
\item Are there multiple mergers going on?

The position of the bent head of Seahorse and the radio relic R3 are in line with the expected positions of the radio relics if the cluster is merging along the North-South direction. But most of the radio emission from Seahorse and R2 are not in line with this picture. The cluster can be merging along two axes one connecting the North-South sub-clusters and the other connecting the Seahorse and R2. As X-ray shocks along both these axes are not detected, unlike the other systems with multiple radio relics like Abell 746 \citep{2023arXiv230901716R} or Toothbrush \citep{2012MNRAS.425L..76B}, the possibility can not be firmly established or ruled out. Also, given the complexity of the radio emission, the radio relic Seahorse might be the product of the off-axis cluster merger and R2 and R3 are products of the independent cluster merger in the Northern subcluster. Though Chandra X-ray temperature maps (Pal et al. in Prep.) are not conclusive, future weak-lensing or deep X-ray maps may shed more light to the merger scenario in the system.

\par
\end{enumerate}
\subsection{The Radio Ring: an `ORC' or a Tailed Source?} We have detected a radio ring of size 100 kpc in the uGMRT 650 MHz image. The radio morphology resembles the newly discovered peculiar radio objects named Odd Radio Circles (ORCs) \citep{2011PASA...28..215N}. To date, all the discovered ORCs have a diameter of 300-500 kpc and are found to be polarized with higher polarization fraction at the edges and gradual depolarization to the centre  \citep{2011PASA...28..215N, 2021PASA...38....3N, 2021MNRAS.505L..11K, 2022MNRAS.513L.101O}. The spectral indices of ORCs are found to be in the range from $-0.8$ to $-1.6$ \citep{2022MNRAS.513.1300N,2023MNRAS.520.1439L,2023ApJ...945...74D}. Three (ORC J2103–6200, ORC J1555+2726, and ORC J0102–2450) out of the discovered ORCs are centred on elliptical galaxies with redshifts of 0.2 to 0.6. \citealp{2023ApJ...945...74D} argued that the ORCs can be a case of galaxy-galaxy mergers. In our case, we did not find any prominent galaxies at the centre of the radio ring. By visually inspecting the morphology, and considering the merging cluster environment, we propose that the radio ring can be formed when the cluster merger shock stirs the ICM and produces enough ram pressure to bend the radio galaxy lobes, similar to the tailed sources, typically seen in cluster mergers. These tailed sources show more polarization in the tails compared to the core \citep{galaxies11030067}. The spectral index of $-0.7 \pm 0.1$, polarization profiles and the environment being a merging cluster hint towards the origin of the radio ring as same as the tailed sources. A very faint galaxy in DSS red images can be traced out at the tip of the ring (Fig.~\ref{ring}) which strengthens the claim of the ring being a tailed source.
\par 
\section{Conclusions} \label{sec:con}
PSZ2 G200.95-28.16 is a low-mass galaxy cluster that hosts three radio relics. We have followed it up with uGMRT at 400 and 650 MHz. The low frequency uGMRT, as well as 1283 MHz archival MeerKAT images combined with X-ray information from Chandra, helped us to study the cluster merger in detail.
\begin{enumerate}[label=(\roman*)]
    \item We have recovered the three radio relics in our low-frequency images. The brightest radio relic Seahorse has a not-so-arc-like morphology that starts with a bent head, then a body with a notch and then an elongated tail. 
    The morphology of radio relic R2 is comprised of two regions---a bright arc-like body with a faint extended arc. The R3 resembles an inverted arc.
    \item The integrated spectral index of Seahorse, R2 and R3 are $-1.1 \pm 0.1$, $-1.4\pm 0.2$, and $-1.6\pm 0.4$. Both the radio relics, Seahorse and R2, show spectral steepening towards the cluster centre. 
    \item Under the assumption of DSA, we deduce the Mach numbers for the shocks responsible for the radio relics are $3.1 \pm 0.8$ (Seahorse) and $2.8 \pm 0.9$(R2).
    \item All the radio relics are polarized at 1283 MHz with the linear polarization fraction being $23 \pm 2\%$, $28 \pm 4\%$ and $58 \pm 9\%$ respectively for Seahorse, R2 and R3. The Seahorse radio relic has two smooth distributions of aligned magnetic field vectors along the two filaments. The magnetic field lines in the R2 are also smoothly aligned with the radio surface brightness. The radio relic Seahorse shows a depolarization trend across the relic width. The magnetic field vectors are not aligned to the radio shock surface for the radio relic R3.
    \item The merging cluster has two sub-clusters with flat surface brightness morphology. We did not detect any shocks towards the directions of the radio relics.    
    \item The radio relics Seahorse and R3 are clear outliers in the mass-luminosity correlation suggested by \citealp{ https://doi.org/10.1093/mnras/stu1658}. We have explored how a low-mass merger can dissipate sufficient energy to a shock if a sufficient mass ratio is available.
    \item We propose possible scenarios of merger geometry given the atypical position of the radio relics compared to the X-ray surface brightness distribution. Given all the evidence, the cluster is most likely an off-axis merger, though the multiple axes merger speculation can not be ruled out given the available data.
    \item We have discovered a radio ring-like source in our high-resolution uGMRT 650 MHz images. We argued that in the merging cluster, the ring-like morphology can be formed by the produced ram-pressure in the ICM by the cluster merger. This source shares the properties of the `Odd Radio Circles' reported recently.
\end{enumerate}
\section{Data Availability}
All the radio and the X-ray data are in public domain and processed data can be accessed from the authors upon reasonable requests. This paper employs a list of Chandra datasets, obtained by the Chandra X-ray Observatory, contained in the Chandra Data Collection (CDC) 304~\dataset[doi:10.25574/cdc.304]{https://doi.org/10.25574/cdc.304}
\section{Acknowledgements}This work makes use of data from three telescopes: uGMRT, MeerKAT and Chandra. We thank the staff of the GMRT that made these observations possible. GMRT is run by the National Centre for Radio Astrophysics of the Tata Institute of Fundamental Research. The MeerKAT telescope is operated by the South African Radio Astronomy Observatory, which is a facility of the National Research Foundation, an agency of the Department of Science and Innovation. QW and DRW acknowledge support from Chandra grant GO7-18117X; this research has made use of data obtained from the Chandra Data Archive and the Chandra Source Catalog, and software provided by the Chandra X-ray Center (CXC) in the application packages CIAO and Sherpa. RK acknowledges the support from the Women Excellence Award grant WEA/2021/000008 from the Science and Engineering Research Board of the Government of India. This research has made use of NASA's Astrophysics Data System Bibliographic Services. This research has made use of the NASA/IPAC Extragalactic Database (NED),
which is operated by the Jet Propulsion Laboratory, California Institute of Technology,
under contract with the National Aeronautics and Space Administration.
\bibliography{sample631}{}
\bibliographystyle{aasjournal}

\end{document}